\documentclass[prb,twocolumn,showpacs]{revtex4}
    \usepackage{epsfig,color}

\begin{document}
    \title{Tuning  Fano resonances by magnetic
    forces for electron transport through  a quantum wire side-coupled to a quantum ring}

    \author{B. Szafran} \affiliation{Faculty of Physics and Applied
    Computer Science, AGH University of Science and Technology, al.
    Mickiewicza 30, 30-059 Krak\'ow, Poland}
    \author{M.R. Poniedzia{\l}ek } \affiliation{Faculty of  Physics and Applied
    Computer Science, AGH University of Science and Technology, al.
    Mickiewicza 30, 30-059 Krak\'ow, Poland}

    \date{\today}

    \begin{abstract}
We consider electron transport in a quantum wire with a side-coupled quantum ring
in a two-dimensional model  that accounts for a finite width of the channels.
We use the finite difference technique to solve the scattering problem as well as
to determine the ring-localized states of the energy continuum.
The backscattering probability exhibits Fano peaks
for magnetic fields for which a ring-localized states appear at the Fermi level. We find that the width of the Fano resonances changes at high magnetic field.
The width is increased (decreased) for resonant states with current circulation that produce the magnetic dipole
moment that is parallel (antiparallel) to the external magnetic field.
We indicate that the opposite behavior of Fano resonances due to localized states with clockwise and counterclockwise currents
results from the magnetic forces which change the strength of their coupling to the channel and modify the lifetime of localized states.

    \end{abstract}
     \maketitle

    \section{Introduction}
Coherent transport properties of mesoscopic and nanoscale conductors is determined by
interference conditions for the electron wave function at the Fermi
level. Quantum dots and rings that are attached by a single contact
to a conducting channel  modify its magneto-transport properties
although they lie outside the classical current path.\cite{webb}
Single subband conductance of the contact is proportional to the
electron transfer probability \cite{lba} and the latter is
particularly sensitive to existence of localized states in the
side-coupled structures. The localized states that belong to the
energy continuum interfere with delocalized states of the channel
leading to an appearance of Fano resonances\cite{fano} in
magnetoconductance when the Fermi level of the channel is degenerate
with the localized energy level. The Fano resonances for quantum
dots and rings connected to a semiconducting channel by one or two
contacts are extensively studied in the context of phase coherence
probes,\cite{clerk} Aharonov-Bohm interferometry,\cite{kprl} Kondo effect,\cite{toriokondo,kang,fang,stefanski}
construction of spin filters,\cite{totio,lee}
 conductance of single-electron transistors,\cite{gores} arrays of quantum dots \cite{zeng,zitko,schmelcher} and artificial defects.\cite{miro,chakrabarti}

The transport properties of singly connected nanostructures are usually studied using theoretical models \cite{zitko,zitkoplusbonca, aligia,chakrabarti,miro,lee,totio,kang,kobayashi} that neglect the finite width of the channels. These models
account for the phase shifts due to the Aharonov-Bohm effect but naturally
overlook the deflection of electron trajectories by classical magnetic forces,
which occurs when the Larmor radius is comparable to the width of the channel.
The purpose of the present paper is to establish
the role of magnetic deflection for the ballistic transport properties of a quantum ring side-coupled to a channel.

The magnetic forces were previously
considered\cite{szafran,szafranpeetersepl} for quantum rings that
are embedded within the channel.\cite{fuhrer,wgw,keyser,muhle,sgs}
In these structures the Lorentz force leads to a preferential
injection of the electron from the input channel to one of the arms
of the ring.\cite{szafran} The preferential injection implies
reduction of the Aharonov-Bohm interference amplitude at the exit to
the output channel.\cite{szafran} A theoretical study of a ring
connected to three channels,\cite{szafranpeetersepl} indicated that
the reduction of the Aharonov-Bohm conductance oscillations due to
magnetic forces is accompanied by a distinct imbalance of the
electron transfer to the two output terminals. This
prediction\cite{szafranpeetersepl} was confirmed in a subsequent
experiment.\cite{strambini} In a recent proposal \cite{ifm,ifm2} of
electronic interaction-free measurement, the idea for the solid
state device employs magnetic forces to deflect the electron
trajectory in a similar manner as beam splitters deflect the photons
in the optical experimental setup,\cite{ifmo} which introduces an
additional interest in magnetic forces.

In this paper we solve the single subband scattering problem for a
quantum ring singly connected to a quantum wire in presence of
perpendicular magnetic field. The resonances that are found in the
electron backscattering probability are confronted with the results
of the stabilization method\cite{stabilization} which allows for
detection of localized states in the energy continuum. We find that
injection of the electron from the channel to the ring is enhanced
at the resonances. The charge distribution within the structure is
not an even function of the magnetic field. This direct effect of
magnetic forces does not influence the linear conductance which is
necessarily an even function of the magnetic field due to the
Onsager symmetry.\cite{microreversibility} We demonstrate that for a
finite width of the channel the Fano resonances of backscattering
probability change in a characteristic manner at high magnetic
field. Typically these resonances appear in pairs. We find that at
high magnetic field one of the resonances widens, and the other
becomes extremely thin. We argue that modification of the width of
resonances is uniquely due to magnetic forces. Thermal stability of
Fano resonances of backscattering probability is also discussed.

The Fano resonances in weak magnetic fields
were considered in particular in Refs. \onlinecite{akis} and \onlinecite{nockel}
for quantum dots embedded within a channel.
Ref. \onlinecite{nockel} describes the quantum-dot-localized states of the energy
continuum between the first and second subband propagation thresholds that are of the odd parity
symmetry with respect to the axis of the system.
These states are
bound, i.e. have an infinite lifetime, since
the leakage to the channel is blocked by the opposite
- even parity - symmetry of the lowest subband of the channel.
The magnetic field breaks the  symmetry with respect to the axis
of the system and allows
for the electron leakage.
The bound states turn into metastable ones with a finite lifetime and the
backscattering probability exhibits Fano peaks
that shift on the energy scale with the magnetic field
according to the sign of the dipole moment produced
by the current circulation in the quantum-dot-localized states.
The study \cite{nockel} is limited to weak magnetic fields, and does
not cover the modification of the lifetime due to magnetic
forces which are the subject of the present paper.

This paper is organized as follows. In Section II we briefly describe the approach applied for the scattering problem.
The stabilization method for localized states detection is sketched in Section III. The results of these two approaches are confronted in Section IV.
Summary and conclusions are given in Section V.

\section{Treatment of scattering problem }

\begin{figure}[ht!]
\begin{tabular}{c}
 \epsfxsize=50mm \epsfbox[119 237 684 726]{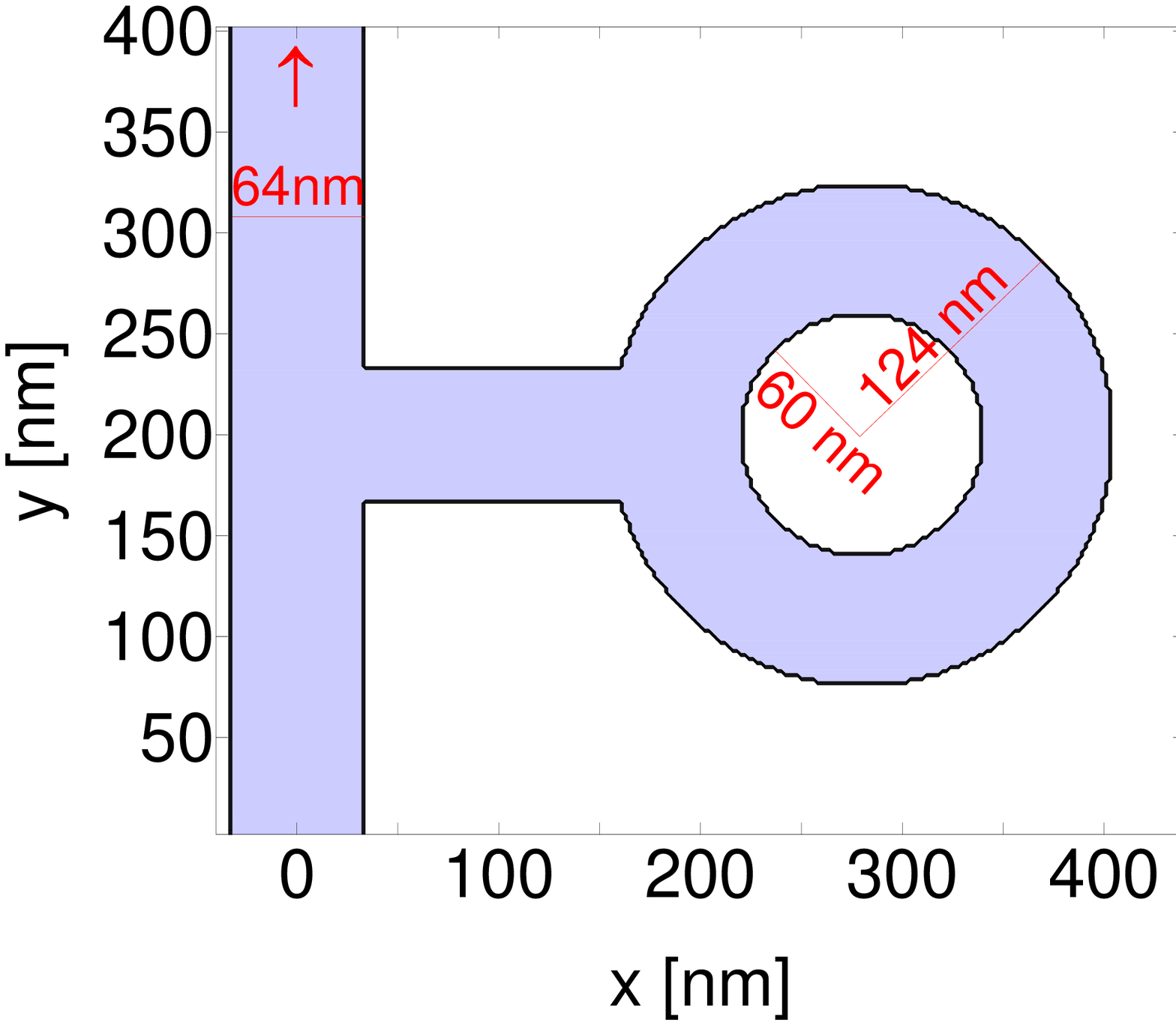} \\
 \epsfxsize=50mm \epsfbox[86 187 634 768]{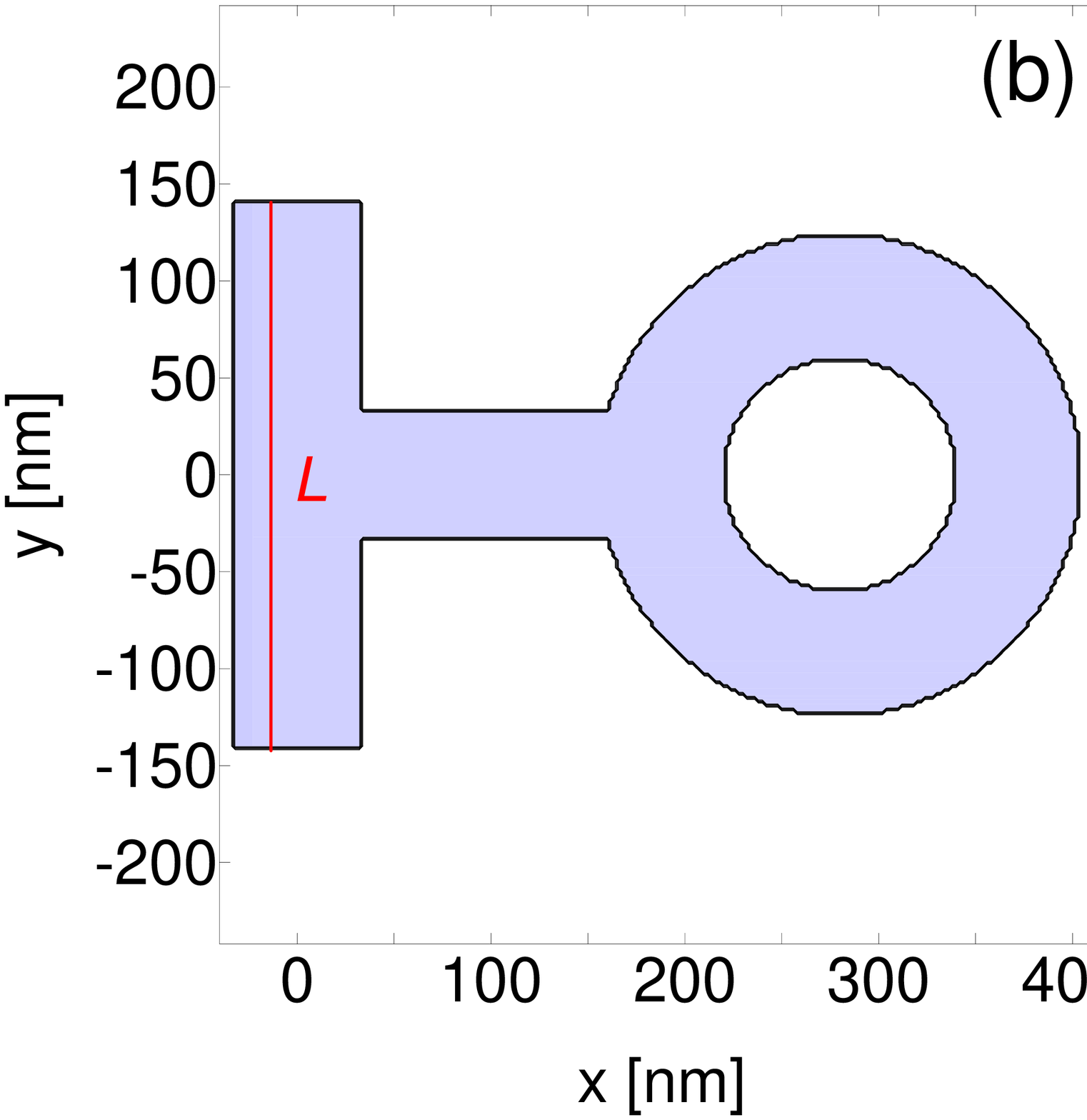}
\end{tabular}
\caption{(a) The system studied in the scattering problem. The
electron is assumed to come from the channel below the contact to
the ring and the vertical channel has an infinite length. (b) The
model system used for determination of the ring-localized states.
$L$ is the finite length of the channel, which varies in the
calculation. The confinement potential is assumed 0 within the blue
area and 200 meV in the outside. }\label{sys}
\end{figure}

    The studied system is presented in Fig. \ref{sys}. We assume that the channels are made of GaAs embedded in Al$_{0.45}$Ga$_{0.55}$As matrix.
    The width of the channels is taken equal to 64 nm, unless explicitly stated otherwise.
    The inner (outer) radius of the ring is 60 nm (124 nm).
    The electron confinement in the growth direction is usually much stronger than the planar one, which justifies application
    of a two-dimensional model that is employed below.
        We consider the Hamiltonian
    \begin{eqnarray}
    H&=&\left({\bf p}+e{\bf A}({\bf
    r})\right)^2/{2m^*}+V(x,y) 
    \label{xx}
    \end{eqnarray}
    where $-e$ is the electron charge ($e>0$) and $m^*=0.067m_0$ is the GaAs electron band effective mass. We apply the Lorentz gauge ${\bf A}=(A_x,A_y,0)=(0,Bx,0)$
    and assume that the confinement potential $V(x,y)$ is zero inside  (the blue area in Fig. \ref{sys}) and $V_0=200$ meV \cite{dodaj}
    outside the channels (the white area in Fig. \ref{sys}).

    We solve the eigenequation
    \begin{equation}H\Psi=E\Psi\end{equation}
     with
    a finite difference approach using a square grid and lattice spacings $\Delta x=\Delta y=2$ nm to determine
 the wave function on a mesh $\Psi_{\mu,\nu}=\Psi(x_\mu,y_\nu)$ . In the calculation we use the gauge-invariant discretization of the kinetic energy operator \cite{governale}
    \begin{eqnarray}
    &&\frac{1}{2m^*}\left({\bf p}+e{\bf A}({\bf
    r})\right)^2\Psi_{\mu,\nu}=\frac{\hbar^2}{2m^*\Delta x^2}\times \nonumber \\&& \left(4\Psi_{\mu,\nu}-C_y \Psi_{\mu,\nu-1}-C^*_y\Psi_{\mu,\nu+1}\right.\nonumber \\ && \left. - \Psi_{\mu-1,\nu}-\Psi_{\mu+1,\nu}\right), \label{new}
    \end{eqnarray}
     with $C_y=\exp\left[-i\frac{e}{\hbar}\Delta x Bx \right]$.

    For description of electron scattering we assume that the input channel is the one below the contact to the ring [Fig. 1(a)]. For the chosen gauge
    the Hamiltonian eigenfunctions in the channel can be written in a separable form
    \begin{equation}
    \Psi(x,y)=\exp(iky)\psi^k(x), \label{psi}
    \end{equation}
    where $k$ is the wave vector and $\psi^k(x)$ is the transverse eigenfunction of the electron in the vertical channel, which is determined
    by solution of a one-dimensional equation obtained by plugging Eq. (\ref{psi}) into the Schroedinger equation (2).
    For the energy that corresponds to the lowest subband the wave function in the incoming lead far from the ring is a superposition of the incident and reflected waves
    \begin{equation}\Psi=\psi^k(x)\exp(iky)+c_{-k}\psi^{-k}(x)\exp(-iky),\label{dirichlet}\end{equation}
    and far in the output lead one has only the transferred wave function
    \begin{equation}\Psi=d_k\psi^k(x)\exp(iky)\label{titi}.\end{equation}
    The values of $c_{-k}$ and $d_k$ are extracted from the finite difference wave functions at the ends of the computational box.

    The boundary condition of the output lead (6) is introduced to the discretized eigenequation (2)
    in a Neumann form
    \begin{equation}
    \Psi_{\mu,\nu+1}=\Psi_{\mu,\nu}\exp(ik\Delta y) \label{bcup}
    \end{equation}
    that results of Eq. (6).
    The boundary condition for the input lead (5) is taken in the Dirichlet form with an
    initial assumption that $c_{-k}=0$. After solution of the algebraic form of the eigenequation we
    extract $c_{-k}$ and $d_k$ by analyzing the finite difference wave function at the ends of the channels. The eigenequation is solved for the new value of $c_{-k}$
    introduced in the boundary condition. The procedure converges after just a few  iterations.\cite{dodaj2}

    After the convergence is reached we evaluate
the backscattering probability by calculating the current fluxes in the incoming channel
    \begin{equation}
    R=\frac{\int dx j^{-k}(x)}{\int dx j^{k}(x)},
    \end{equation}
     where $j^k$ is the probability density current associated to the incoming part of the
wave function of Eq. (\ref{dirichlet}),
    \begin{equation}
    j^{k}(x)=\frac{\hbar }{m^*}|\psi^k(x)|^2(\hbar k+eBx) \label{ji}
    \end{equation}
    and $j^{-k}$ corresponds to the backscattered one
    \begin{equation}
    j^{-k}(x)=\frac{\hbar }{m^*}|c_{-k}|^2|\psi^{-k}(x)|^2(-\hbar k+eBx).
    \end{equation}
    The backscattering probability is related to magnetoconductance \cite{lba} as $G(B)=\frac{2e^2}{h}\left[1-R(B)\right].$

     The magnetic forces modify the width of the Fano resonances of the backscattering probability not only as a function of the magnetic field but
     also as a function of the energy. In finite temperature a transport window is opened near the Fermi level $E_F$ and the narrow Fano resonances
     are likely to be thermally unstable. The stability of Fano resonances against thermal excitations is below
     estimated by the linear response formula
    \begin{equation}
    \overline{R}(E_F)=\int R(E) \left(-\frac{\partial f}{\partial E}\right) dE,\label{tildet}
    \end{equation}
    where $f$ is the Fermi function
    $f=\left(e^{(E-E_F)/k_BT}+1\right)^{-1}$.
    $T$ stands  for the temperature and $k_B$ for the Boltzmann constant.

\begin{figure}[ht!]

\epsfysize=65mm \epsfbox[30 253 688 807]{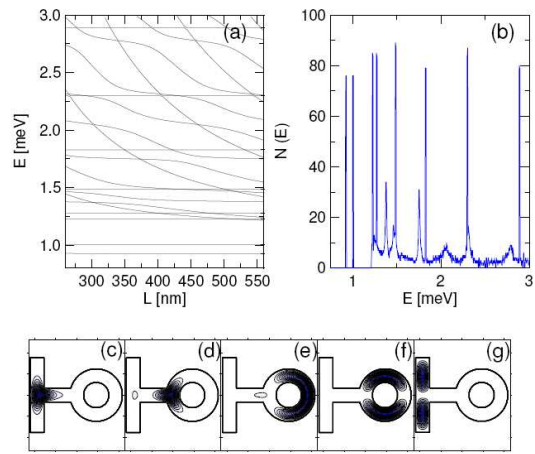} 


\caption{(a) Energy spectrum for the channel of a finite length
($L$) for $B=0$. (b) The peaks indicate the energies of the
localized states extracted of the energy spectrum according to  Eq.
(12). (c-e) Charge densities of the five lowest energy eigenstates
for $L=352$ nm. Plot (c) corresponds to the ground-state, (d) to the
first excited state etc.}\label{st}
\end{figure}

\section{Detection of localized states}

     In order to determine the resonant states localized in the ring we use the stabilization method. \cite{stabilization}
     For this purpose we assume that the vertical channel has a finite length [see Fig. \ref{sys}(b)].
     Then we solve the eigenequation (4) with boundary conditions  that require the wave functions to vanish
     at all the edges of the computational box. We calculate the energy spectrum as a function of the length
     of the channel $L$. The energy spectrum that is discrete for finite $L$ is displayed in Fig. \ref{st}(a) for $B=0$.
     We see that some energy levels
          decrease as $L$ grows. They correspond to electron states in the channel.
           The energy of  states localized in the ring are independent of
           $L$.

           The two lowest energy levels of Fig. 2(a) correspond to states localized at one of T-junctions \cite{bqd} present in the system -- one between the main vertical channel  and the short horizontal one [Fig. 2(c)], and the other between the latter and the quantum ring [Fig. 2(d)]. These two states are not only localized but energetically bound - their energies (0.925 meV and 1.006 meV) lie below the continuum threshold. The threshold is determined by the bottom of the lowest subband  ($k=0$) for the infinite vertical channel. For 64 nm - wide channel the energy continuum
           starts above 1.147 meV ($B=0$).
            The energy levels of Fig. 2(a) above the continuum threshold that are independent of $L$ correspond to
           the ring-localized states, see Figs. 2(e) and 2(f). The fourth excited state [Fig. 2(g)] found for $L=352$ nm
           corresponds to the electron within the channel,
            and its energy distinctly depends on $L$ [see Fig. 2(a) for $L=352$ nm].

     The spectral positions of
     the localized states are determined by counting the states of the energy close to $E$,
     \begin{equation}
     N(E)=\int dL \sum_l \delta(|E-E_l(E)|;dE),
     \end{equation}
     where $l$ numbers the Hamiltonian eigenvalues for the finite size of the system, and $\delta(|E-E_l(L)|;dE)$
     is equal to 1 for $|E-E_l(L)|<dE$ and 0 otherwise. In Fig. \ref{st}(b) we plotted $N(E)$ calculated
     for the energy spectrum of Fig. \ref{st}(a)
     with the energy window $dE=5\mu$eV.

\begin{figure}[ht!]

\epsfysize=65mm \epsfbox[80 242 536 817]{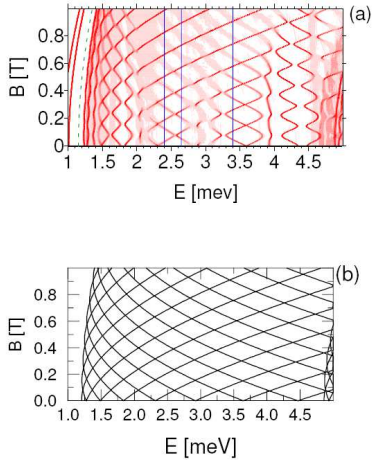} 

\caption{(a) Positions of localized states in the energy
continuum as calculated by formula (12). The darker the shade of red
the larger value of $N(E)$. The thin vertical lines indicate
energies of $2.4$, $2.65$ and $3.4$ meV that are considered in
detail below. The green dashed line indicates the continuum threshold (ground-state energy of the electron
within the channel for $k=0$).
(b) Energy spectrum of the closed circular quantum
ring that is not connected to the channel. }\label{steb}
\end{figure}

\begin{figure}[ht!]
 \epsfxsize=85mm \epsfbox[81 181 573 610]{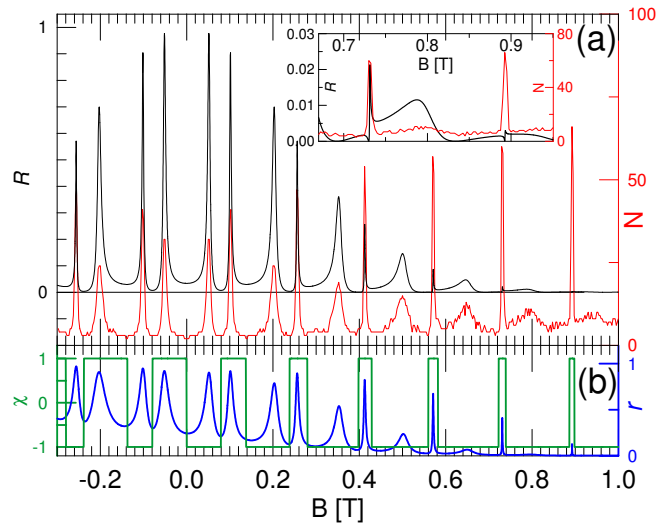}
\caption{(a) Black line shows the electron backscattering probability for energy $E=2.65$ meV and the red one -- the value
of the resonance detection counter $N(E)$ [Eq.(12)]. The inset shows a zoom of high $B$ part of the figure.
(b) The value of $\chi$ plotted in green indicates the direction
of the current circulation inside the ring: $\chi=1$ (-1) corresponds to counterclockwise (clockwise) direction.
The blue line shows the fraction of the probability density $r$ that is contained within the ring, i.e. for $x>32.5$ nm of
the computational box of Fig. \ref{sys}(a).
}\label{e265}
\end{figure}


\begin{figure}[ht!]
\epsfysize=65 mm \epsfbox[31 256 389 821]{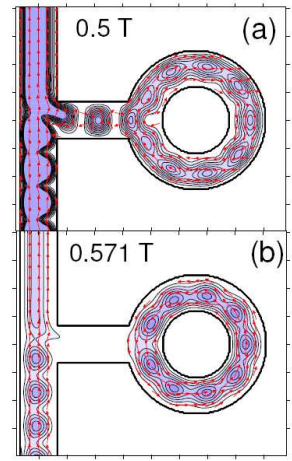}
\caption{The contours show the charge density for two peaks of backscattering probability of Fig. \ref{e265} obtained for $E=2.65$ meV at $B=0.5$ T (a) and at $B=0.751$ T (b). The arrows show the probability current distribution.
}\label{ew265}
\end{figure}

Figure \ref{steb} shows how the maxima of $N(E)$ shift with $B$
 [Fig. \ref{steb}(a)] as compared to the energy
spectrum of a closed \cite{cqr} quantum ring [Fig. \ref{steb}(b)]
that is not connected to the channel. The two lowest-energy lines of
Fig. \ref{steb}(a) correspond to states localized at the junctions.
The resonances that are higher in the energy correspond to states
localized in the ring. The positions of resonances oscillate with
magnetic field with a period of 0.165 T that results from the
Aharonov-Bohm effect for a ring of an effective radius of 92 nm.
 The resonances enter into avoided crossings that result from angular momentum mixing of closed-ring states.
 The mixing is due to the presence of the
contact that breaks the rotational symmetry and is the strongest at odd multiples of half of the flux quantum threading the ring. Above 4.6 meV the
states of the second subband - with wave functions that change
sign across the channel appear in Fig. \ref{steb}(a). For the closed
ring these states appear near 4.9 meV [Fig. \ref{steb}(b)].

\begin{figure}[ht!]
\begin{tabular}{c}
 \epsfxsize=85mm \epsfbox[57 590 577 820]{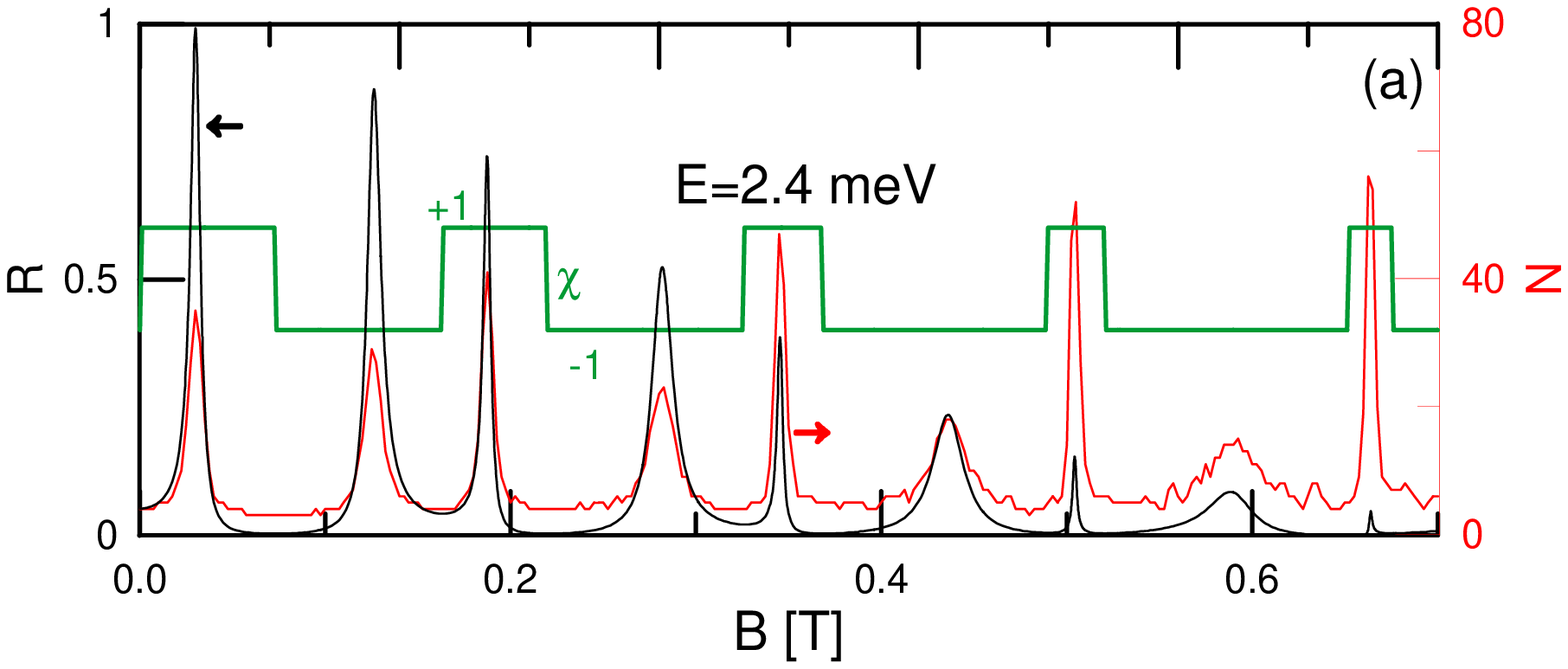} \\
 \epsfxsize=85mm \epsfbox[57 590 577 820]{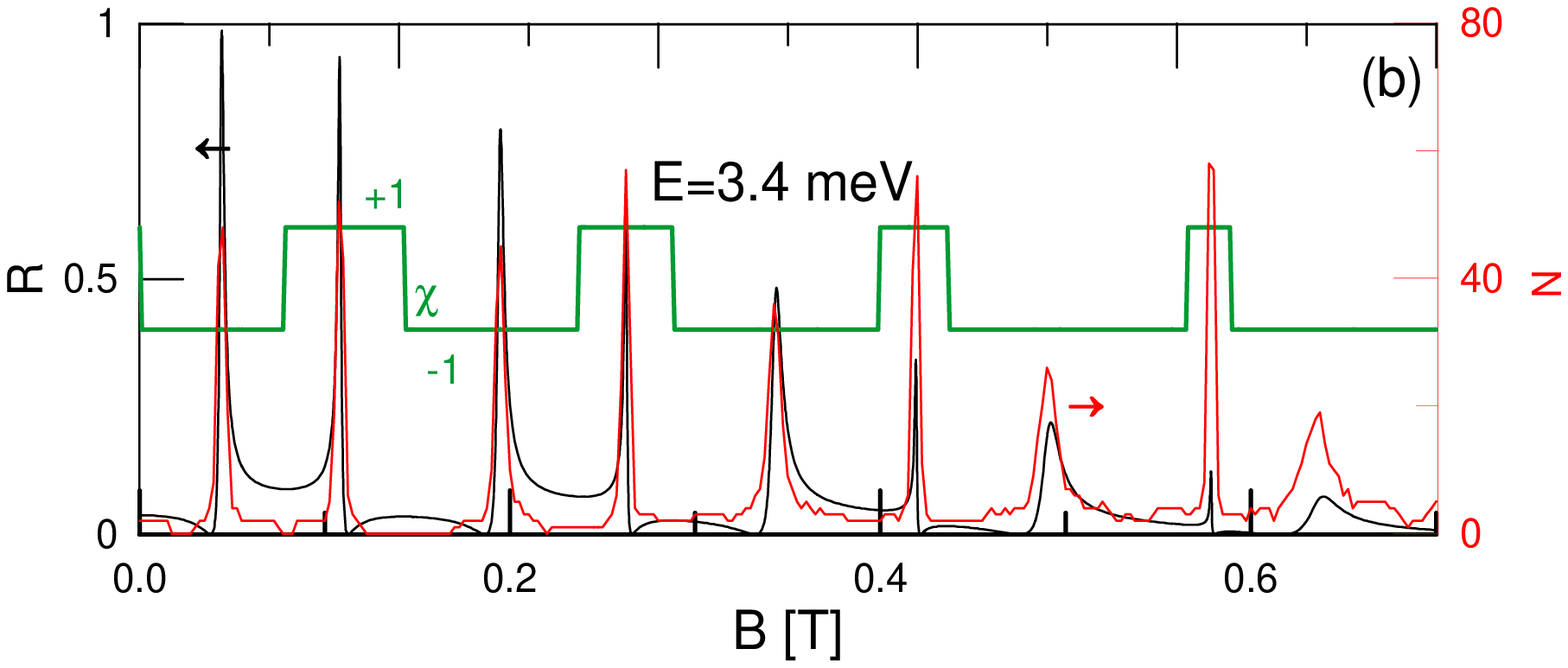} \end{tabular}

\caption{Backscattering probability (black line, left axis), the
resonance counter [Eq.(12), red line, right axis] and the direction
of the current circulation within the ring in the scattering
eigenstates [green line, $\chi=+1$ (-1) corresponds to
counterclockwise (clockwise) orientation]. Plots (a) and (b) were
prepared for $E=2.4$ meV and 3.4 meV, respectively. }\label{inne}
\end{figure}

\begin{figure}[ht!]
 \epsfxsize=65mm \epsfbox[65 255 557 615]{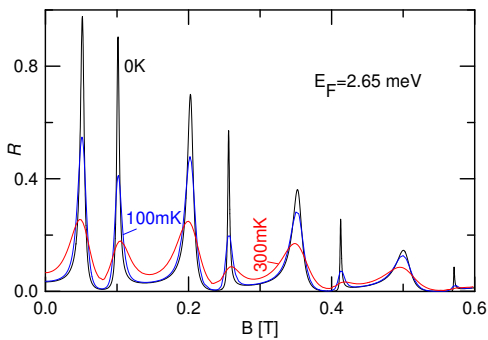}

\caption{The backscattering probability averaged according to the
linear response formula (11) for the Fermi energy assumed at 2.65
meV for 0K (black line), 100 mK (blue line) and 300 mK (red line).
}\label{inneuj}
\end{figure}

\begin{figure}[ht!]
 \epsfxsize=70mm \epsfbox[47 300 768 814]{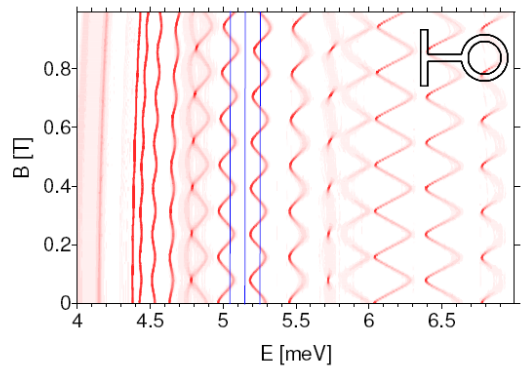}
\caption{Same as Fig. 4(a) but for the channel width reduced to 32 nm.
The inner and outer radii of the side-coupled ring 76 and 108 nm, respectively. The
geometry is displayed in the inset.
The thin vertical lines indicate
energies of $5.05$, $5.16$ and $5.26$ meV that are considered in
Fig. 9.  }\label{32}
\end{figure}

\begin{figure}[ht!]
\begin{tabular}{c}
 \epsfxsize=65mm \epsfbox[157 355 661 717]{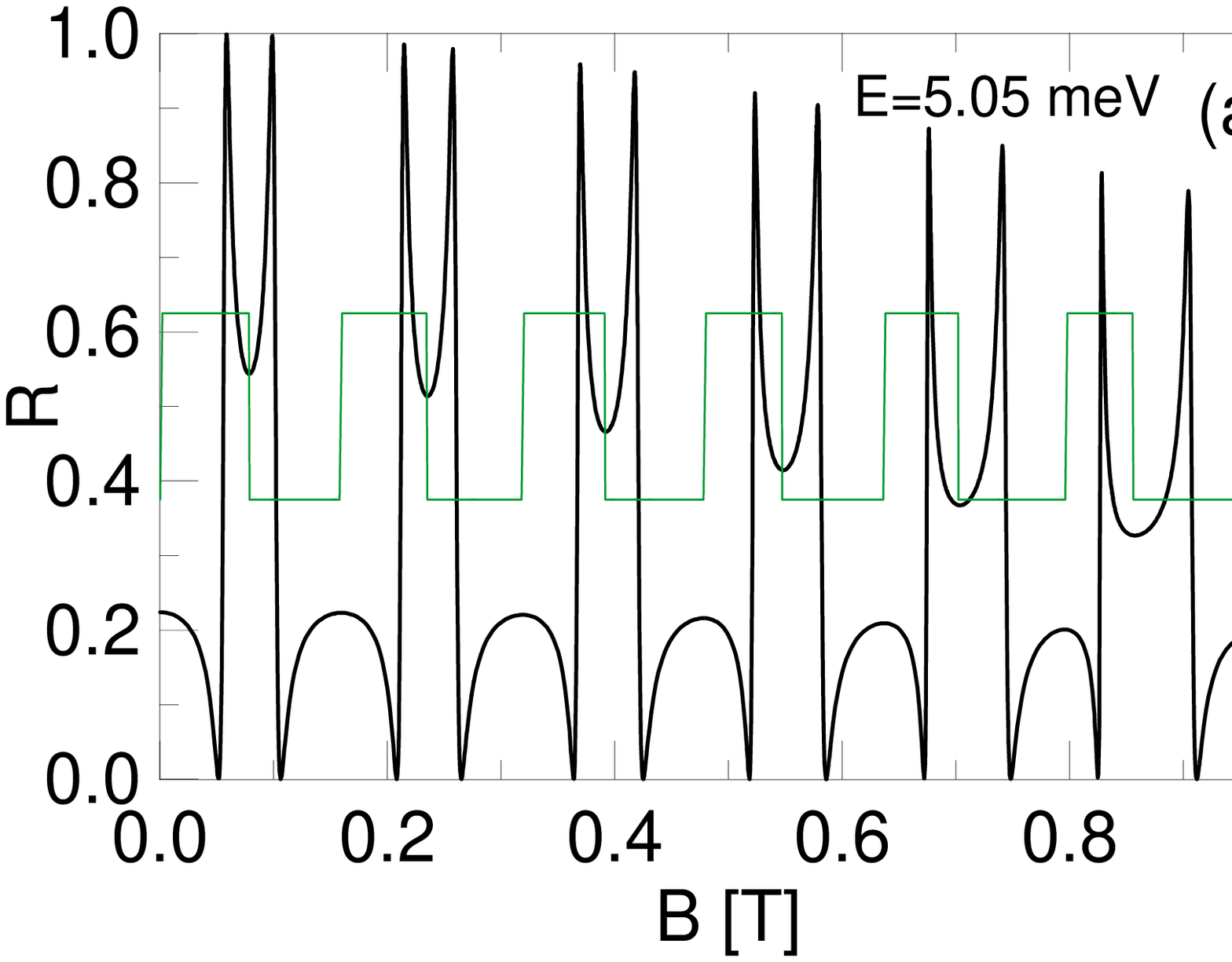} \\
  \epsfxsize=65mm \epsfbox[157 355 661 717]{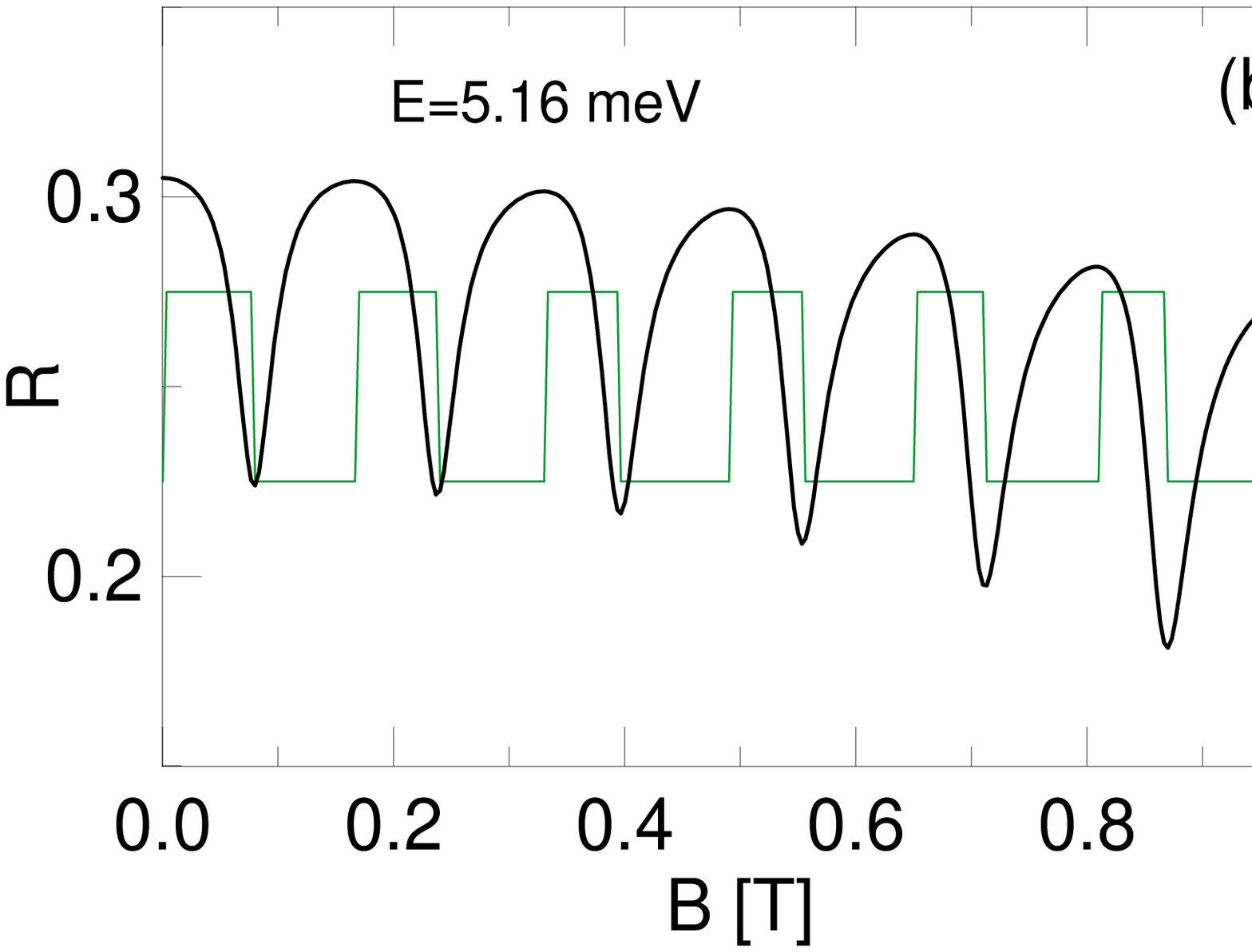} \\
   \epsfxsize=65mm \epsfbox[157 355 661 717]{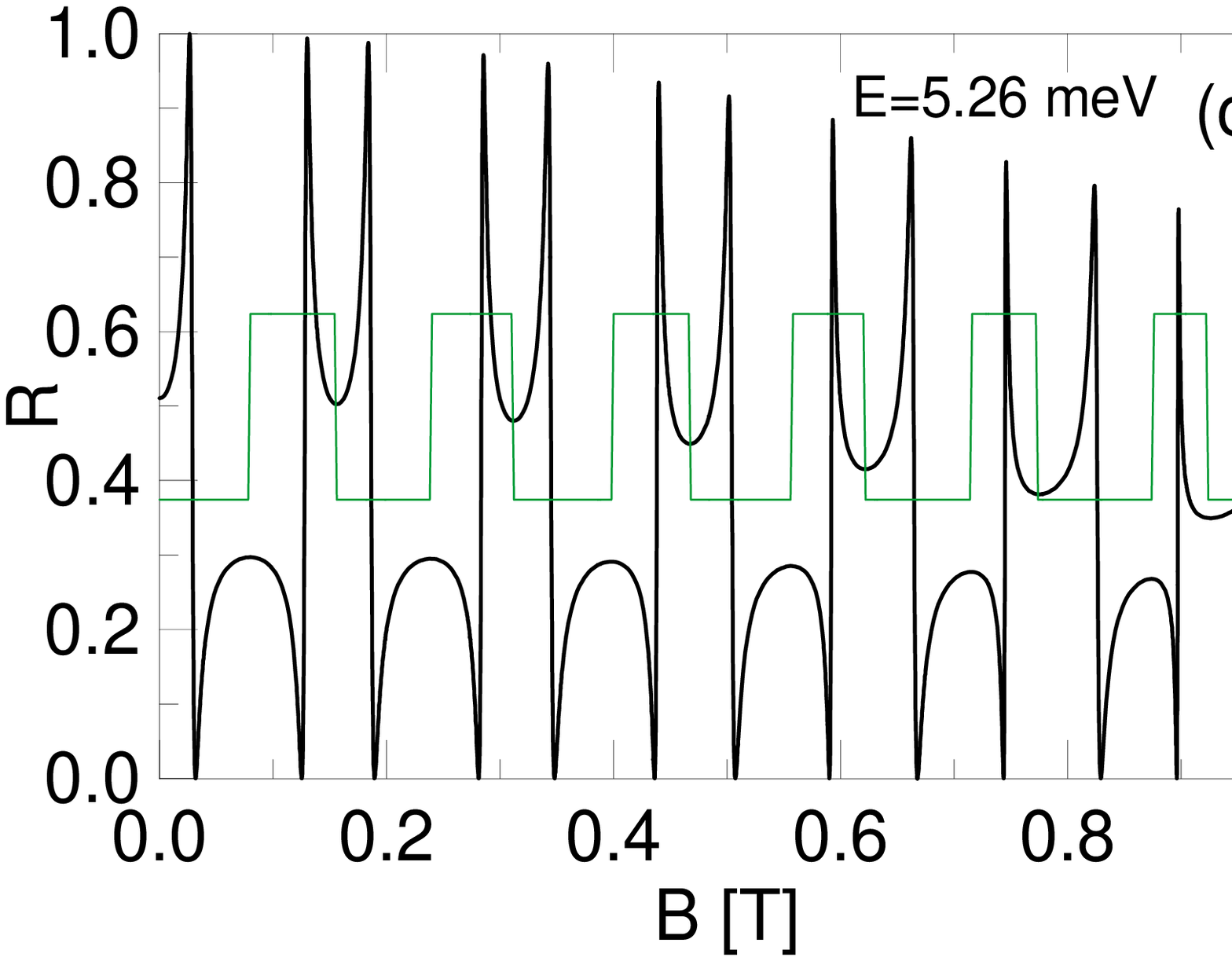}
 \end{tabular}

\caption{Backscattering probability (black line, left axis), and the direction
of the current circulation within the ring in the scattering
eigenstates [green line, $\chi=+1$ (-1) corresponds to
counterclockwise (clockwise) orientation]. Plots (a,b,c) were
prepared for $E=5.05$, 5.16 and 5.26 meV, for the structure parameters of  Fig. 8. }\label{inne}
\end{figure}

\section{Resonances in the quantum resistance}
Now let us look at the electron backscattering probability that is plotted with the black line in Fig. \ref{e265}(a) for $E=2.65$ meV.
The red lines in Fig. \ref{e265}(a) show the resonance detection counter $N(E)$ of Eq. (12).
$R$ exhibits peaks as a function of the magnetic field. These peeks perfectly coincide on the magnetic field scale with the positions
of the ring-localized states as determined by the stabilization method.
The backscattering occurs only when the channel state carrying
the current is degenerate with a localized state within the ring. Outside the degeneracy
the structure is nearly transparent for the electron transfer.
The peaks of $R$ have distinctly asymmetric shape which is a characteristic signature of a resonance involving a localized state of the energy continuum.\cite{fano}

In Fig. \ref{e265}(b) we plotted with the blue line the fraction $r$ of the probability density stored by the computational box of Fig. \ref{sys}(a) that is contained within the ring (outside the vertical channel in fact).
We notice that the electron enters the ring for magnetic fields for which a localized state is present at the considered energy.
Note that $R$ is an even function of $B$ but $r$ is not [$r(B)\neq r(-B)$]. For $B>0$ the Lorentz force deflects the electron trajectories to the left
of its momentum vector, hence the penetration of the wave function to the ring is hampered for positive and enhanced for negative magnetic field.

The height of $R$ maxima is reduced for high magnetic field. For
$B>0$ the Lorentz force pushes the electron to the left edge of the
vertical channel. The electron with wave function shifted to the
left edge of the channel seems not to notice the presence of the
ring and passes through the contact with a nearly 100\% probability.
At high $B$ the presence of the ring still produces the Fano peaks
but on a tiny scale [see the inset to Fig. 4(a)]. In this sense the
Lorentz force for $B>0$ assists in the electron transport across the
contact. Note, that due to the microreversibility
relation\cite{microreversibility} we have $R(B)=R(-B)$ although for
negative $B$ the electron penetration to the ring is enhanced.

The results for  $N(E)$ as presented in Fig. \ref{e265}(a) are cross section of Fig. \ref{steb}(a) taken along the constant energy line.
 We can see that the line for $E=2.65$ meV crosses the resonances
of Fig. \ref{steb}(a) that grow or decrease in energy as the magnetic field grows.
The magnetic dipole moment generated by currents is $\mu=-\frac{dE}{dB}$. For a strictly one dimensional
closed circular quantum ring  $\mu=-\frac{1}{2}e {\bf r} \times {\bf j}$, where ${\bf j}$ stands for the probability density current.
Thus, the resonances that grow (decrease) in the energy correspond
to localized states in which the current circulates counterclockwise (clockwise) around the ring and produce magnetic dipole
moment that is antiparallel (parallel) to the external magnetic field. The direction of the current flow within the ring as found in the solution of the scattering problem is denoted by $\chi=\pm 1$ [green line in Fig. 4(b)], where the plus sign stands for the counterclockwise
circulation.  Note, that $\chi$ changes sign between each minimum of the transfer probability. For $B>0$ the
resonances that increase (decrease) in width corresponds to clockwise (counterclockwise) current circulation. At high positive magnetic field
the current circulation within the ring in the scattering eigenstates is counterclockwise with the exception of $B$ ranges that surround
the sharp Fano resonances [Fig. 4(b)].

Figure \ref{ew265} shows the charge density and current distribution
for two backscattering  peaks of Fig. \ref{e265}(a), a wide one [Fig.
\ref{ew265}(a)] and a narrow one [Fig. \ref{ew265}(b)].  In both
these cases the charge density is pushed to the left with respect to
the electron ''velocity'', i.e. to the external edge of the ring for
the wide peak [Fig. \ref{ew265}(a)] and to the internal edge of the
ring for the narrow one [Fig. \ref{ew265}(b)]. The clockwise
orientation of the current circulation is at high magnetic field
translated by the Lorentz force into stronger coupling of the
localized states to the conducting channel [Fig. \ref{ew265}(a)].
The electron of the ring-localized state is ''ejected'' into the
horizontal link to the main channel by the Lorentz force. The charge
density within the horizontal channel acquires similar values to the
ones within the ring. The stronger coupling for these states leads
to an increased width of the resonance, or in other words
to a decreased lifetime of the ring localized state.
Opposite effects are
observed for resonances with localized states of counterclockwise
current circulation, for which the Lorentz force tends to keep the electron within the ring,
increases the localized state lifetime and thus reduces the width of the resonance in the transfer probability.

Let us count the resonances for positive $B$
starting from zero field, for the energies
marked by thin blue lines in Fig. 3(a). For $E=2.65$ meV -- the case that was
discussed above -- the Fano resonances that become narrow at high $B$
correspond to {\it even} resonance numbers. For  $E=2.4$ meV the first
resonance is obtained for a localized state with a counterclockwise
current circulation that grows in the energy for increasing $B$ [Fig. 3(a)].
Accordingly, Fig. 6(a) shows that the resonances that become narrow at high $B$
correspond to {\it odd} resonance numbers [see Fig. 3(a)]. For $E=3.4$ meV [Fig.
\ref{inne}(b)] the odd (even) peaks increase (decrease) in width as
for $E=2.65$ meV in consistence with the order of resonances crossed
at this energy in function of $B$ of Fig. 3(a).

Fig. 7 shows the backscattering probability averaged over the
transport window opened near the Fermi energy assumed equal 2.65 meV
for $T=100$ mK and 300 mK according to Eq.(11) . The resonances that
corresponds to ring localized states with counterclockwise current
circulation are less thermally stable than the ones with the
clockwise flowing current.

For completeness we present results for the structure that is closer
to the one-dimensional limit. For this purpose we reduce the channel
width from 64 to 32 nm. The shape of the ring with thin channels is
displayed in the inset to Fig. 8. We keep the axis of the vertical
channel as well as the center of the ring in unchanged positions,
and set the outer and inner radii of the ring equal to 108 and 76
nm, respectively to maintain the same Aharonov-Bohm period as in the
structure considered above.

Fig. 8 shows the positions of localized states as determined by the
stabilization method. The stronger transverse confinement increases
significantly the energy of the low part of the spectrum and
enlarges the avoided crossing due to the presence of the contact.
The vertical lines in Fig. 8 indicate the energies for which we
display also the 0K backscattering probability: 5.05 meV [Fig.
9(a)], 5.16 meV [Fig. 9(b)] and 5.26 meV [Fig. 9(c)].   The energy
of 5.05 meV lies within the wide gap opened by the avoided crossing
of localized energy levels. The case presented in Fig. 9(b)
corresponds to the energy for which no resonances are found in
function of the magnetic field. The oscillations of $R$ have a small
amplitude and are due to the Aharonov-Bohm phase shifts within the
ring. For the two other energies considered here, the Fano
resonances are distinctly present in $R(B)$ dependence [Fig. 9(a)
and 9(c)]. The maxima of $R$ are close to 1 and decrease in function
of $B$ more slowly than for the wide structure considered above [see
Figs. 4 and 6], since the shift of the electron density within the
channel due to the Lorentz force is hampered by the much stronger
confinement. The Fano resonances that become thin in both cases
considered in Figs. 9(a) and 9(c) correspond to counterclockwise
circulation of the current within the ring. Concluding, the
modification of the width and height of resonances are of the same
origin and form as for the wide channels, only the effects of
magnetic forces are of a reduced strength. We also note that for the
thin channels the backscattering probability is of the order of
0.25\% outside the resonances  as compared to nearly zero for the
wider channels discussed above.

\section{Summary and Conclusion}
We have studied the electron transport in a channel side-coupled to
a quantum ring taking into account the finite width of the
structure. The evaluated backscattering probability exhibits Fano peaks
for magnetic fields for which the ring-localized states of the energy continuum
are degenerate with the Fermi level.  We have demonstrated that the ring-localized
states that correspond to current circulation
producing a magnetic dipole moment that is antiparallel to the external magnetic field
produce very thin Fano resonances. The
localized states with the dipole moment that is aligned with the magnetic field vector result in resonances
that become wider at high magnetic field. The latter are also more
thermally stable. The opposite behavior of these two types of
resonances is due to shifts of the radial wave function of
ring-localized states induced by magnetic forces which increase or
decrease the coupling to the channel depending on the direction of
the current circulation. The described effect can be used to tune
the Fano resonances for potential
applications\cite{totio,lee,schmelcher} or for magnetic forces
detection.

    {\bf Acknowledgements}
This work was performed within a research project N N202 103938
supported by Ministry of Science an Higher Education (MNiSW) for 2010-2013.
 Calculations were performed in
    ACK\---CY\-F\-RO\-NET\---AGH on the RackServer Zeus.


\begin{thebibliography}{00}
\bibitem{dodaj} The barrier height $V=200$ meV is large as compared to the energy range of the
    lowest subband transport which for the system of Fig. \ref{sys} corresponds to the energy $E$ lower than 4.5 meV.
    The wave functions vanish exponentially in the barriers with a penetration length equal to $\hbar/\sqrt{2m^* (V-E)}\simeq \hbar/\sqrt{2m^* (V)}$,
    which equals roughly 1.7 nm. This is small as compared to the width of the channels (64 nm).
    The effect of the finite barrier is an increased effective width of the channels. The bridge in the structure of Fig. \ref{sys}
    is necessary to allow the channel electrons to reach the ring.
\end{thebibliography}
    \end{document}